\documentclass{article}

\usepackage{PRIMEarxiv}

\usepackage[utf8]{inputenc} % allow utf-8 input
\usepackage[T1]{fontenc}    % use 8-bit T1 fonts
\usepackage{hyperref}       % hyperlinks
\usepackage{url}            % simple URL typesetting
\usepackage{booktabs}       % professional-quality tables
\usepackage{amsfonts}       % blackboard math symbols
\usepackage{nicefrac}       % compact symbols for 1/2, etc.
\usepackage{microtype}      % microtypography
\usepackage{lipsum}
\usepackage{fancyhdr}       % header
\usepackage{graphicx}       % graphics
\usepackage{steinmetz}
\usepackage{amsmath}
\usepackage{amssymb}
\usepackage{svg}
\usepackage{float}
\graphicspath{ {./figures/} }
\title{Thermodynamics-inspired Macroscopic States of Bounded Swarms}

\author{
  Hossein Haeri, Kshitij Jerath \\
  Department of Mechanical Engineering \\
  University of Massachusetts Lowell \\
  Lowell, MA\\
  \texttt{\{hossein\_haeri, kshitij\_jerath\}@uml.edu} \\
   \And
  Jacob Leachman \\
  School of Mechanical and Materials Engineering \\
  Washington State University \\
  Pullman, WA\\
  \texttt{jacob.leachman@wsu.edu} \\
}

\begin{document}
\maketitle

% \begin{abstract}
% \lipsum[1]
% \end{abstract}

% keywords can be removed
% \keywords{First keyword \and Second keyword \and More}

% \section{Introduction}
% \lipsum[2]
% \lipsum[3]

\begin{abstract}
The collective behavior of swarms is extremely difficult to estimate or predict, even when the local agent rules are known and simple. The presented work seeks to leverage the similarities between fluids and swarm systems to generate a thermodynamics-inspired characterization of the collective behavior of robotic swarms. While prior works have borrowed tools from fluid dynamics to design swarming behaviors, they have usually avoided the task of generating a fluids-inspired macroscopic state (or macrostate) description of the swarm. This work will bridge the gap by seeking to answer the following question: is it possible to generate a small set of thermodynamics-inspired macroscopic properties that may later be used to quantify \textit{all} possible collective behaviors of swarm systems? In this paper, we present three macroscopic properties analogous to pressure, temperature, and density of a gas, to describe the behavior of a swarm that is governed by only attractive and repulsive agent interactions. These properties are made to satisfy an equation similar to the ideal gas law, and also generalized to satisfy the virial equation of state for real gases. Finally, we investigate how swarm specifications such as density and average agent velocity affect the system macrostate.

\end{abstract}

% \IEEEpeerreviewmaketitle

\section{Introduction}
Recently, there has been considerable interest in better understanding collective behavior in a diverse range of natural as well as engineered complex systems \cite{Werfel754}\cite{naldi2010mathematical}.
Swarms can be found in both the natural and engineered worlds, and represent complex systems in which agent-based interactions are often simple and local.
Like other self-organizing complex systems, swarms are characterized by the formation of one or more macroscopic-scale collective patterns which can be extremely difficult to predict. These difficulties are in fact observed across a diverse range of self-organizing multi-agent systems (MAS), including traffic jams \cite{Jerath2015}, economic systems \cite{Bajo2017}, and even human societies \cite{diamond1997guns}. Quantification and classification of the collective behavior of a swarming system not only is necessary to better understand and predict the dynamics of these complex systems, but also it can give us more intuition about more complex self-organizing systems \cite{Moussaid2011}.

%     \begin{figure}
%     \includegraphics[width=\linewidth]{figures/birds_medium_resolution.jpg}
%     \caption{`Swarming' behavior of a flock of seagulls on the Caspian Sea beach, Gilan, Iran (Hossein Haeri, 2017). A self-organizing multi-agent system may exhibit several different kinds of collective behavior or patterns, and a single macroscopic-scale  description may be beneficial for understanding them.
%     }
%     \label{fig:fish}
%     \end{figure}

% Recent works have also attempted to leverage the apparent similarities between the collective behaviors of fluid particles and swarming agents to obtain greater insights into swarm dynamics and control \cite{Spears2009} \cite{Spears2006}.

\begin{figure}[ht]
    \begin{center}
    \includegraphics[width=0.7\linewidth]{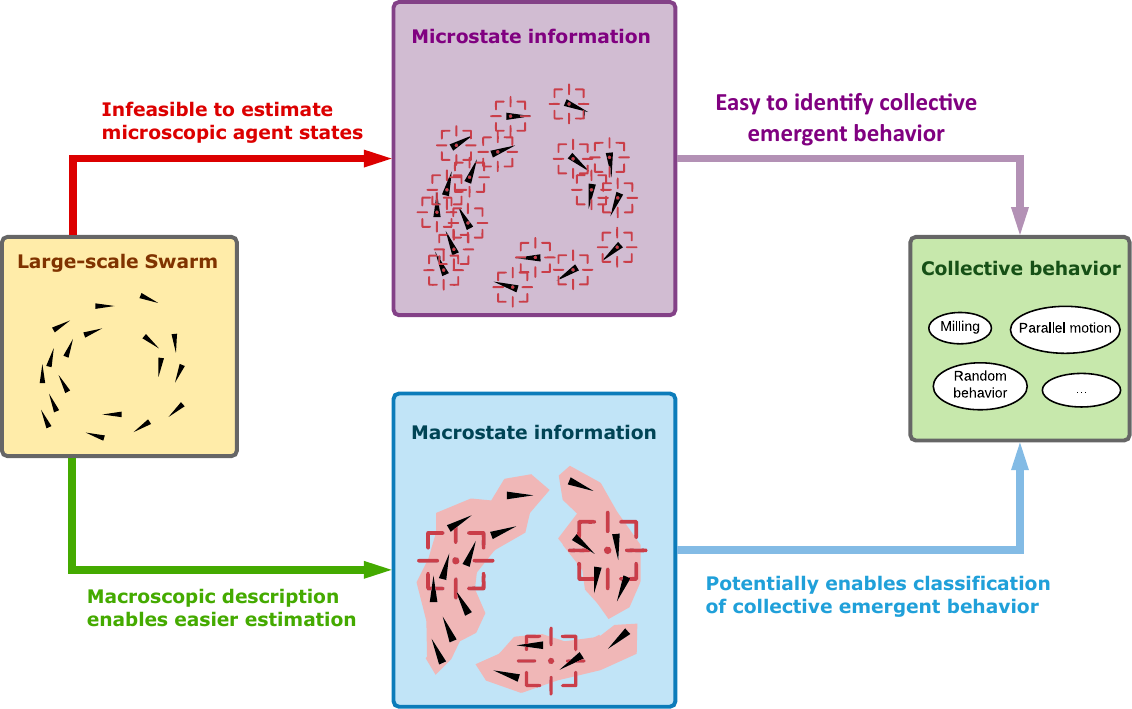}
    \caption{Emergent behaviors in a swarm can be easily categorized if the swarm's microstate information (i.e. states of all agents) is accessible, but this is infeasible from a practical viewpoint. A thermodynamics-inspired macroscopic description may enable easier state estimation and classification of collective, emergent behavior, especially for unknown swarms.}
    \label{fig:chart}
    \end{center}
    \end{figure}

Generally, this task can be conducted in two ways as shown in Fig. \ref{fig:chart}. The first method relies on obtaining the microstate information, i.e. the state information of all the individual agents in the swarm, and then using it to classify collective behavior. This approach is infeasible in practice due to the considerable difficulties in obtaining microstate information. The second method relies on estimating the macrostate information using macroscopic-scale state variables and then classifying collective behaviors. This approach offers greater potential, but requires significant investigation beyond what exists in the state-of-the-art.

Similarly, researchers in the field of thermodynamics have long known about the Lennard-Jones potential function \cite{Lennard-Jones1924}, which is used to characterize the interactions between pairs of fluid particles. However, knowledge of classifying `collective' behaviors of materials into different phases (e.g., solid, liquid, and gas) predates the work of Lennard-Jones in 1924. These observations drive us to a natural query: can swarms be modeled as a thermodynamic system? 

\begin{figure}[t]
    \centering
    \includegraphics[width=0.5\linewidth]{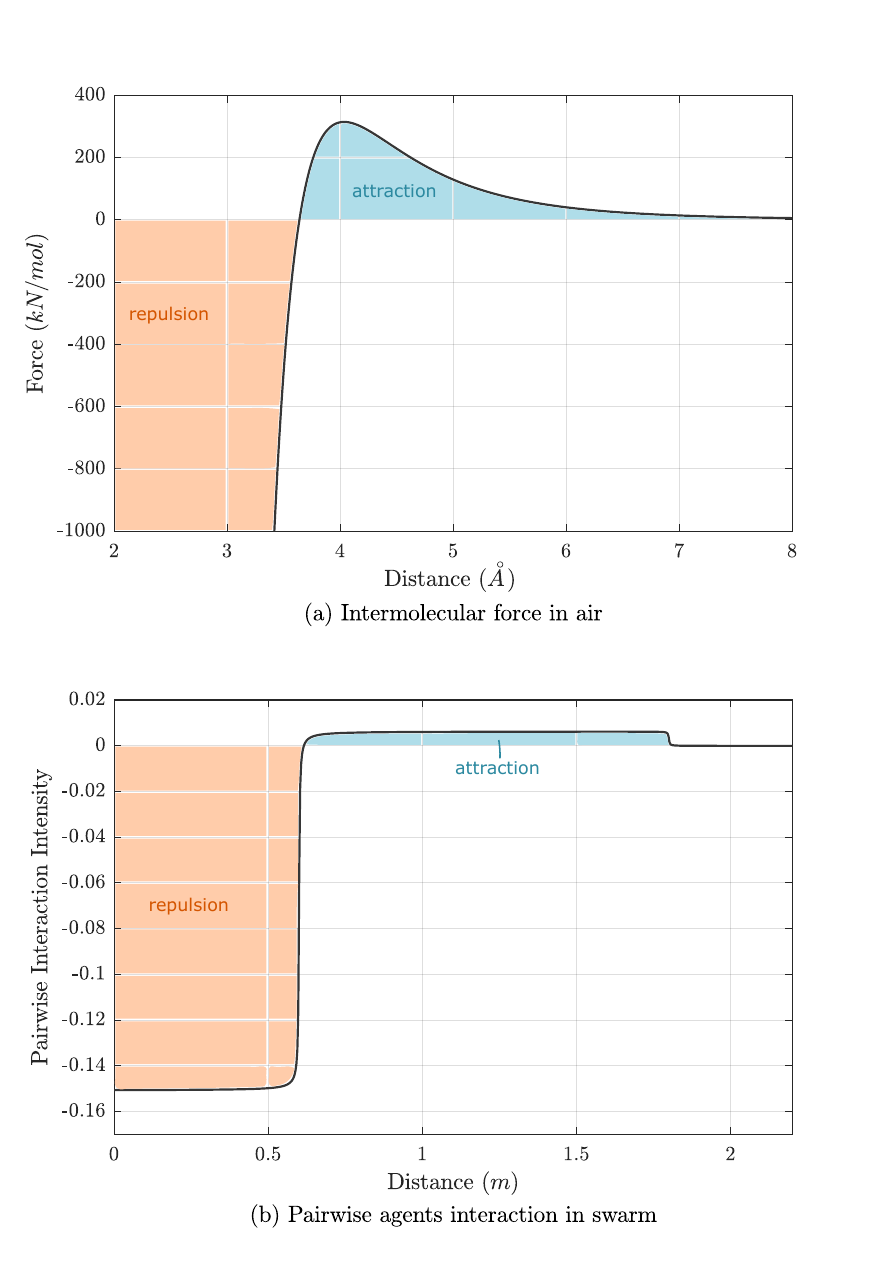}
    \caption{(a) Lennard-Jones force function for air. Differentiating Lennard-Jones potential function with respect to intermolecular distance ($d$)  gives the intermolecular force function: $F_{mol}(d)=\frac{\partial \epsilon}{\partial d}$ where $\epsilon=4(\epsilon_0)[(\frac{d}{r})^{12} - (\frac{d}{r})^6)]$, $\epsilon$ is intermolecular potential energy, $r$ represents intermolecular relaxed distance, and $\epsilon_0$ is a constant \cite{Friedman1957} (b) Thermodynamics-inspired pairwise agent interaction function $f_{pair}$.}%
    \label{fig:lennard}%
\end{figure}

There are several analogies that can be drawn between fluid and swarm systems. First, like swarming agents, fluid particles typically interact locally with each other but not over long distances. This is evident from Fig. \ref{fig:lennard}(a) which shows the asymptote for the Lennard-Jones potential tending to zero with increasing distance between fluid particles. Second, like fluid molecules in Lennard-Jones function, most agent interactions are usually modeled as being attractive and/or repulsive in nature (in addition to other local responses) \cite{Gazi2003}\cite{Romanczuk2012}\cite{parrish1997animal}. 

In fact, fluid-like behavior has been reproduced in robotic swarms for collective movement \cite{yamagishi2017collective} and surveillance tasks \cite{Spears2009} \cite{Spears2006}, but the authors do not yet know of any literature that explicitly defines macroscopic swarm states in the context of fluids. Existing literature provides only limited insights into the macroscopic-scale description of swarm properties \cite{VanDykeParunak:2001:ESM:375735.376024}. For example, Jantz et al. \cite{Jantz1997} have used thermodynamics-related concepts to indirectly measure the performance of a swarm assigned the task of finding an exit using statistical interactions between agents. However, these works have not been extended to generate a macroscopic description of a swarm system.

Nonetheless, there are some recent efforts \cite{porfiri2016effective} \cite{mwaffo2015collective} to to better understand the order-disorder phase transition in a simple class of swarms based on Vicsek model \cite{Vicsek1995}. Our work follows the similar  The authors believe that defining the swarm's macrostate could play a key role in quantifying, classifying and ultimately predicting the collective, emergent behavior of swarms in the future.

% For example, Couzin \cite{Couzin2002} postulated that collective behavior of swarms could be replicated if each agent possessed three zones: one each corresponding to repulsion, orientation and attraction. The subsequent swarming dynamics could be classified on the basis of the magnitudes of zone radii for each of the three zones \cite{Couzin2002}.

In the following sections, first, a general definition of swarm macrostate is described, then, a swarming model that is inspired by the works of  Gazi et al. \cite{Gazi2003} and Couzin et al. \cite{Couzin2002} is introduced. Next, two macroscopic properties, viz. swarm temperature ($T_s$) and swarm pressure ($P_s$) are defined analogously to their thermodynamic counterparts. The swarm pressure and temperature are shown to satisfy the appropriate ideal gas law. We then extend these ideas to the virial equation of state and derive virial coefficients for swarms that are analogous to real gas-like behaviors.

\section{Swarm Macrostate}
A swarm macrostate can potentially encapsulate global-scale information about the system and help classify collective behavior and more easily. At the microscopic scale, the state of a two dimensional swarm consists of the positions $(x_i,y_i)$ and velocities $(\dot{x}_i, \dot{y}_i)$ of the $N$ agents in the system, and hence the state-space dimension is 4$N$. Within this 4$N$-dimensional state-space, the macrostate may be defined as a set of macroscopic-scale variables ($\underline{\phi}$) whose evolution is restricted to a low-dimensional manifold such that $h(\underline{\phi}) = 0$. For example, the macroscopic state variables of a gas (i.e. pressure, volume, and temperature) are related via the low-dimensional manifold given by $h(P, V, T) = PV - nRT = 0$, which represents the ideal gas law. Now the obvious questions are: what are these macroscopic properties or macrostate variables need to describe a swarm system and how should we find them? In the following sections, we propose thermodynamics-inspired swarm macrostate variables ($\underline{\phi}$) and seek to find the nonlinear low-dimensional manifold $h(\underline{\phi}) = 0$.

\section{Swarm System Dynamics}
To introduce the concept of a fluid-like swarm macrostate, the swarm dynamics will be restricted to two spatial dimensions. The collective behavior of the swarm will evolve in this 2-D world, with the state $\underline{x}_i$ of each agent given by:

    \begin{equation}
    \underline{x}_i = [x_i, y_i, \theta_i]^T
    \label{eq:x}
    \end{equation} 

\noindent where $[x_i, y_i]$ denotes the agent's  position  vector, and $\theta_i$ is its heading. The single integrator dynamics of each agent are given by the following equation:

    \begin{equation}
    \underline{\dot{x}}_i =
        \begin{bmatrix}
            \dot{x}_i \\
            \dot{y}_i \\
            \dot{\theta}_i 
        \end{bmatrix}
        = 
        \begin{bmatrix}
            v sin(\theta_i) \\
            v cos(\theta_i) \\
            k(\angle{}\vec{f_i} - \theta_i)
        \end{bmatrix}
    \label{equ:x_dot}
    \end{equation}

\noindent where $v$ denotes the constant linear speed for all agents, $\angle\vec{f}_i$ represents the heading of the swarm interaction vector $\vec{f}_i$ for agent $i$, which will be discussed in subsection \ref{SubSec:Attracion-repulsion-model}, and the constant $k$ is proportional to $|\vec{f}_i|$.

Many self-organizing complex systems, regardless of their domain of origin, exhibit a local behavior which is known as "short-range activation and long-range inhibition" \cite{Willis2016}. In swarm systems, this agent-based behavior is modeled as attractive and repulsive interactions. In this notion, each agent attempts to avoid straying away from the  group (long-range inhibition), and is also actively repelled by neighboring agents (short-range activation). The choice of the specific functions that determine the attractive and repulsive behaviors of individual agents can vary significantly, and this depends heavily on the application being modeled. For example, while biologists use certain behavioral functions to model natural biological systems \cite{Topaz2004}\cite{Couzin2002}, roboticists and engineers may design different agent-based interaction to help the swarm succeed at a particular task \cite{Brambilla2012}\cite{Gazi2004}\cite{Yu2013}\cite{Xue2010}. In fact, the similar behavior was proposed for fluid molecules where square well potential was utilized to determine the relationship between potential parameters and the macroscopic behavior of the fluid \cite{Chrisman1973}. While the included work presents a specific model for attractive and repulsive interactions between agents, the insights from this study can potentially be extended beyond these functional forms. In the next few sections we make an analogy between swarm agents and fluid particles for a specific functional form.

\subsection{Attractive/Repulsive Interaction Model}
\label{SubSec:Attracion-repulsion-model}
In the following discussions we define a swarm interaction intensity $f$ and the swarm interaction vector $\vec{f}$ to model the attractive and repulsive interactions between neighboring agents. The swarm interaction intensity $f$ belongs to the class $\mathbb{C}^1$, i.e. its first derivative is continuous. The attraction/repulsion function is modeled along the lines of the Lennard-Jones potential function. Consequently, the interaction strength between pairs of agents is assumed to decay to zero at large distances to reflect the local nature of these interactions \cite{Brambilla2012}. However, unlike the Lennard-Jones potential function, the interaction function is assumed to be bounded below and above to generate realistic swarm dynamics. Fig. \ref{fig:lennard} shows the similarities and differences between the proposed interaction function and the Lennard-Jones force function \cite{Lennard-Jones1924}.

The pair-wise agent interaction intensity is evaluated by summing the attraction $(f_a)$ and repulsion $(f_r)$ intensities as a function of the distance $d$, as follows:
\begin{equation}
f_{pair}(d) = f_r(d) + f_a(d)
\label{eq:f_pair}
\end{equation}
where,
\begin{align}
    f_r(d) &= - \frac{k_r}{\pi} \left\{\frac{\pi}{2} - tan^{-1}(s_r(d-r_r))\right\} \\
    f_a(d) &= \frac{k_a}{\pi} \left\{\frac{\pi}{2} - tan^{-1}(s_a(d-r_a))\right\}
\end{align}
and $k_a$ and $k_r$ are the unitless repulsion and attraction `gains' which represent the magnitude of attraction and repulsion of an agent by another agent. In most swarming scenarios, the repulsive behavior is expected to dominate at shorter distances, so the repulsion gain is modeled to be higher than the attraction gain (i.e. $k_r > k_a$). Additionally, the parameters $r_r$ and $r_a$ specify the radii of repulsion and attraction zones, respectively. Moreover, the parameters $s_r$ and $s_a$ characterize how quickly the repulsive and attractive interaction intensities fade with distance. The authors would like to note that the interaction intensities $f_r$ and $f_a$ represent scalar quantities and must not be interpreted as forces between agent pairs.

    \begin{figure}[b]
    \begin{center}
    \includegraphics[width=0.4\linewidth]{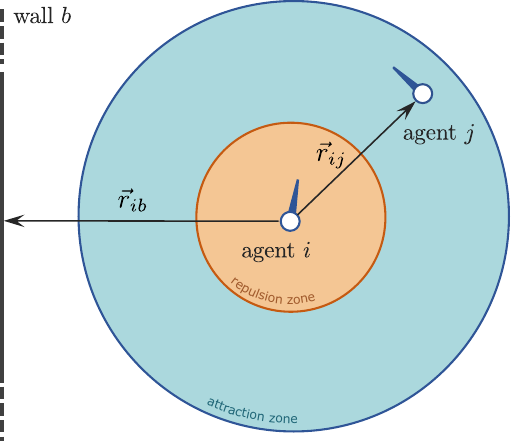}
    \caption{Schematic showing interactions between agent $i$ and its neighbouring agent $j$, as well as between agent $i$ and wall boundary constraint. The distances between them dictate whether the agent $i$ experiences repulsion or attractive effects.}
    \label{fig:geometry}
    \end{center}
    \end{figure}

\subsection{Modeling Repulsive Boundary Constraints}
\begin{figure}[b]
    \begin{center}
    \includegraphics[width=0.4\linewidth]{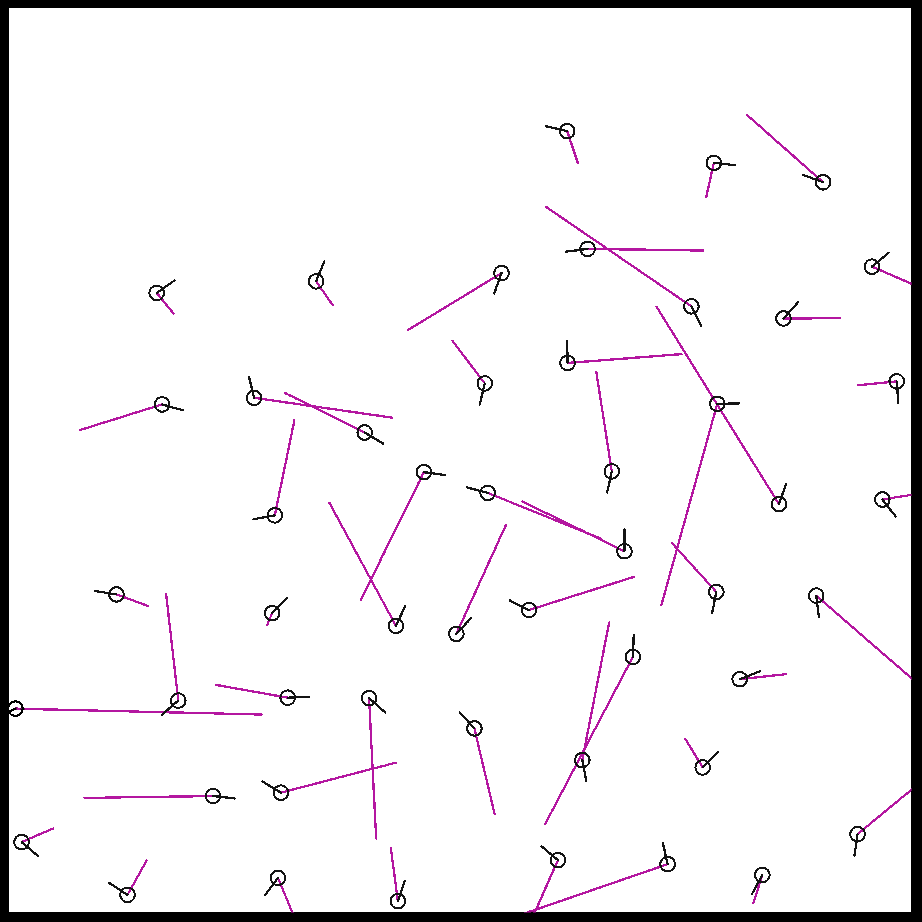}
    \caption{Two-dimensional simulator snapshot. The short black lines represent the headings and the purple lines represent the interaction vectors $\vec{f}_i$. The swarming area is a $5 m \times 5 m$ square (i.e. $V=25$) and is fixed. In this study, the other agent specifications are: $k_r=0.15$, $k_a=0.006$, $s_a=s_r=5000$, $r_r = 0.6$, and $r_a=1.8$. Also, for all repulsive boundaries: $k_b=0.1$ and $s_b$=500.
    }
    \label{fig:sim}
    \end{center}
    \end{figure}

Most thermodynamic studies of fluids are typically conducted within a set of physical constraints which define the system boundaries. Similarly, swarming operations such as surveillance or search-and-rescue may require the agents to be restricted to a specified area. To replicate these operational conditions, we define a set of virtual boundary constraints around the swarming system which repel agents in their vicinity. As with the pairwise repulsion intensity, we define the boundary repulsion intensity as:
\begin{equation}
    f_{bound}(d') = -k_b\left\{\frac{\pi}{2} - tan^{-1}(s_b d')\right\}
\label{eq:f_bound}
\end{equation}
where $d'$ represents the distance of the agent in question to a boundary constraint, and $s_b$ represents how quickly the boundary effects fade with distance. As shown in Fig. \ref{fig:sim}, the value of $s_b$ is chosen to model rapid decay of repulsive boundary effects. %Fig. \ref{fig:geometry} denotes the inter-agent and agent-boundary interactions in the simulated system.

Eventually, the dynamics of an individual swarm agent are a combination of the pairwise interactions with all other swarm agents, as well as the repulsive boundary effects. These are quantified as a swarm interaction vector $\vec{f}_i$ for agent $i$, which is defined as follows:
\begin{equation}
\vec{f_i} = \sum\limits_{j=1 , i \neq j}^{N} \frac{\vec{r}_{ij}}{|\vec{r}_{ij}|} f_{pair}(|\vec{r}_{ij}|) +
\sum\limits_{b=1}^{B} \frac{\vec{r}_{ib}}{|\vec{r}_{ib}|} f_{bound}(|\vec{r}_{ib}|)    
\label{eq:f_i}
\end{equation}
where $\vec{r}_{ij}$ represents the position vector from agent $i$ to agent $j$, and $\vec{r}_{ib}$ is normal distance of agent $i$ from boundary constraint $b$. These notations are visualized alongside the shaded repulsion and attraction zones for agent $i$ and neighboring agent $j$ in Figure \ref{fig:geometry}. Additionally, the parameters $N$ and $B$ represent the number of agents and boundary constraints (or walls), respectively.

The direction of the swarm interaction vector dictates the desired heading of the agent, and its magnitude $|\vec{f_i}|$ specifies the intensity of the resultant swarm interaction. These quantities are used to simulate the single integrator swarm dynamics as discussed in (\ref{equ:x_dot}). Figure \ref{fig:sim} shows a representative snapshot of the simulation environment with thick black boundary constraints, agents as circles, and the swarm interaction vectors for each agent as purple lines. 

\section{Thermodynamic Analogy for Swarms with Non-interacting Agents ($f_{pair}=0$)}
In thermodynamics, the ideal gas law is an equation relating pressure ($P$), temperature ($T$) and volume ($V$) of an ideal gas as follows: $PV = nRT$, where $n$ represents the number of moles, and $R$ is the universal gas constant. Unlike a real gas, an ideal gas assumes the lack of interactions between gas molecules. The universal gas constant can be written as $ R = N k_b /n$, where $N$ is number of the particles and $k_b$ is Boltzmann's constant, resulting in the following representation of the ideal gas equation:
\begin{equation}
    \dfrac{PV}{NT} = k_b
\label{eq:ideal_gas_law}
\end{equation}
which indicates that the value of $PV/NT$ remains constant for an ideal (or perfect) gas. A key query now presents itself: Does there exist an analogous `thermodynamic law' that describes a swarm system at the macroscopic scale, given that agent behaviors may potentially be governed by non-physical laws? Specifically, can we define swarm macro-properties that are analogous to pressure and temperature, i.e. swarm pressure $(P_s)$ and swarm temperature $(T_s)$, such that they satisfy following equation:
\begin{equation}
    \dfrac{P_sV}{NT_s} = k_s
\label{eq:ideal_swarm_EOS}
\end{equation}
where $V$ represents the volume constraints placed on the swarm, $N$ represents the number of swarming agents, and $k_s$ may be considered to be a `swarm constant' analogous to Boltzmann's constant. We now define swarm pressure and swarm temperature by drawing inspiration from analogous definitions for fluid systems. While the analysis assumes ideal gas behavior and neglects pairwise agent interactions (i.e. $f_{pair}=0$), this assumption is relaxed in later sections. The reader should also note that due to the boundary constraints placed on the swarm in this study, only isochoric (i.e. constant volume) thermodynamic processes can occur in the simulated system.

\subsection{Swarm Pressure ($P_s$)}
From a macroscopic perspective, the pressure of a gas is simply the magnitude of force applied on the wall per unit area. On the other hand, from a microscopic perspective, the kinetic theory assumes that pressure is caused by the force associated with individual atoms striking the walls. However, in the context of swarms, agents typically do not apply any `force' to their surrounding environment. Therefore, we build upon the concept of gaseous pressure, and evaluate swarm pressure by measuring boundary effects using the boundary repulsion intensity $f_{bound}$. To measure these boundary effects in the two-dimensional simulator, we evaluate time-averaged magnitude of boundary interactions across the entire agent population divided by total length of system boundaries $L$: 
 \begin{equation}
 P_s = \frac{1}{(t-t_0)L} \int_{t_0}^{t} \sum_{i=1}^{N} |f_{bound,i}|dt
 \label{eq:P_s}
 \end{equation}

\subsection{Swarm Temperature ($T_s$)}
\begin{figure}
    \begin{center}
    \includegraphics[width=.5\linewidth]{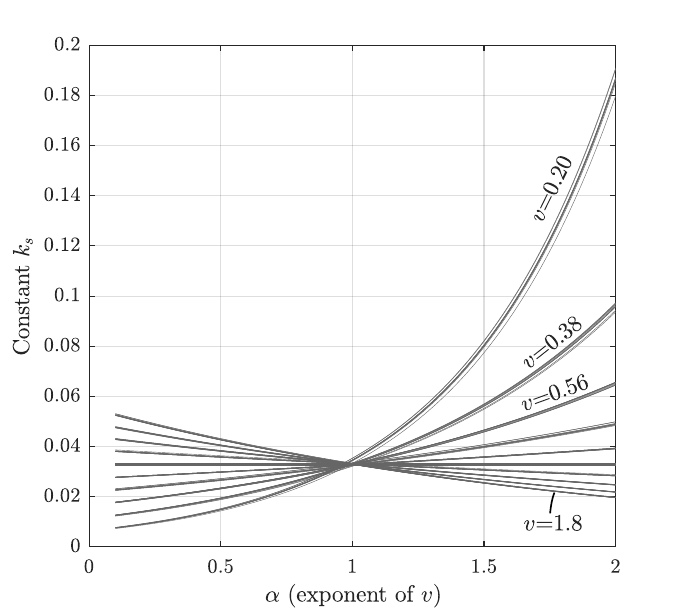}
    \caption{To ensure that (\ref{eq:ideal_swarm_EOS}) is satisfied, $\alpha$ is chosen so as to result in an identical value of constant $k_s$ across all simulations. Simulations are performed 2,500 times for 50 different agent velocities ($0.2<v<1.8$) and 50 different swarm densities ($2<N/V<5.6$) to evaluate the constant $k_s = P_sV/NT_s$. In this figure, each line bundle corresponds to a specific $v$ value, while the different lines within a bundle correspond to various swarm densities $(N/V)$. The plot indicates that $k_s$ is relatively independent of swarm density $N/V$, and that there is only one value of $\alpha$ for which $k_s$ remains constant across the entire parameter set.}

    \label{fig:alpha_plot}
    \end{center}
    \end{figure}

In thermodynamics, finding an explicit microscopic-scale expression for temperature is a difficult task. The gas kinetic theory, however, does indicate that temperature is directly proportional to the total kinetic energy of the system which itself is directly proportional to square of average velocities of particles. However, since the swarm agent interactions are typically not physics-based, the expectation that temperature is proportional to the square of agent velocities may not be justified. To examine the relationship between swarm temperature ($T$) and the absolute agent velocity ($v$), we run the simulation for various agent numbers $N$ and velocities $v$ and record corresponding $P_s$ values. Defining the swarm temperature $T_s \triangleq v^{\alpha}$, we evaluate the term $P_s V/N T_s (=k_s)$ for various values of $\alpha \in [0.1, 2]$. Figure \ref{fig:alpha_plot} shows the variability in the `constant' $k_s$ for various possible exponents of $v$. It is evident that $k_s$ does not necessarily remain constant for any arbitrarily chosen definition of $T_s$. As is shown, $\alpha \approx 1$ results in $k_s$ being almost constant (in this case $k_s \approx 0.303$) and  $T_s = v$ satisfies the ideal equation of states quite well. While these simulations help us define the relationship between the swarm temperature and agent velocity, they do not yet include pairwise agent interactions. As a result, these discussions do not yet help us determine a realistic macroscopic-scale description of the swarm. The next section relaxes this assumption by including pairwise agent interactions and modeling the swarm as a real gas.
 
\section{Empirical Equation of State for Swarms with Interacting Agents ($f_{pair} \neq 0$)}

 Our previous discussions indicated that we can identify a thermodynamic-inspired macroscopic-scale description of a swarm based on the ideal gas law. Specifically, we demonstrated that in an `ideal gas-like' swarm, i.e. swarming without agent-to-agent interactions, the macroscopic-scale state variables are related such that the system evolves on a low-dimensional manifold characterized by (\ref{eq:ideal_swarm_EOS}). In this section, we relax the `ideal gas' assumption to model realistic swarm behavior with agent-to-agent interactions given by (\ref{eq:f_pair}). In thermodynamics, the value of any macroscopic property $X$ can be written as the sum of an ideal term and a residual term \cite{Assael1973}, where the significance of the residual term may attributed to non-idealities such as particle interactions:
\begin{equation}
X = X_{ideal} + X_{residual}
\label{eq:ideal+residual}
\end{equation}

%To that extent, swarm's macrostate defined as a 3D point in the macro-properties space. We showed in an 'ideal swarm', i.e. swarming without agent-to-agent interactions, all emergent macrostates are laid on a lower dimensional manifold characterized by equation \ref{eq:ideal_swarm_EOS}. In This section, we generalize previous manifold for the swarm with agent-to-agent interaction, where the interaction is determined by equation \ref{eq:f_pair}. In thermodynamics, the value of any property X can be written as the sum of an ideal term and a residual term \cite{Assael1973}, where any significance of the residual term is attributed to non-idealities, i.e. particles interactions:
%\begin{equation}
%X = X_{ideal} + X_{residual}
%\label{eq:ideal+residual}
%\end{equation}

The empirical equation of state, also known as the virial equation of state, is an equation which generalizes the ideal gas law to real gases, as follows:
\begin{equation}
\dfrac{P}{\rho R T} = 1 + B\rho + C\rho^2 + D\rho^3 + \: \dots
\label{eq:virial_EOS}
\end{equation}
where $\rho = N/V$ denotes the density, the density-dependent terms on the right-hand side represent `residual' terms, and the constants $B$, $C$, $D$, etc. are functions of temperature and depend on the fluid substance being modeled. Specifically, the second virial coefficient, $B$, arises from the interaction between a pair of molecules, the third virial coefficient, $C$, depends upon interactions in a cluster of three molecules, $D$ involves a cluster of four molecules, and so on for the following higher-order terms \cite{Assael1973}. 

Taking inspiration from the virial equation of state for gases, we can generalize the ideal swarming equation of state to model realistic swarms by adding  residual terms which correct the deviations caused by agent interactions:
\begin{equation}
\frac{P_s V}{N k_s T_s} = 1 + a_1\left(\frac{N}{V}\right) + a_2\left(\frac{N}{V}\right)^2 + ... + a_m\left(\frac{N}{V}\right)^m
\label{eq:swarm_EOS}
\end{equation}
\noindent where $a_1$, $a_2$, etc. are functions of the swarm temperature $T_s$. Theoretically, an infinite number of residual terms could be used to model real gases in (\ref{eq:swarm_EOS}), resulting in an infinite number of virial coefficients. However, the magnitude of virial coefficients for higher order terms is much smaller than for lower-order terms, since the likelihood of simultaneous higher-order interactions between several agents drops significantly. Consequently, we restrict the analysis to four higher order terms. The exact number of higher-order virial terms that may be neglected perhaps cannot be generalized for arbitrary swarms, just like it cannot be generalized for arbitrary gases.
%While the presented relation cannot be generalized in the context of arbitrary swarms, we can it at least gives us the intuition of expecting a decreasing tail of virial terms.

The virial coefficients are identified by performing Monte Carlo simulations with $p$ different agent velocities and $q$ different agent densities ($\rho = N/V$). The swarm pressure $P_s$ given by (\ref{eq:P_s}) and swarm temperature $T_s$ are recorded for each simulation, and are used to calculate the cumulative residual term $\psi = (P_s V/N T_s k_s)-1$. As expected, the agent non-idealities result in different magnitudes of $\psi$ for each simulation run. Using a swarm density matrix $\Gamma$ for $q$ different values of densities,  
\begin{equation}  
\Gamma=
\begin{bmatrix}
    \rho_{1}       & \rho_{1}^2  & \dots &  \rho_{1}^m \\
    \rho_{2}       & \rho_{2}^2  & \dots &  \rho_{2}^m \\
    \vdots & \vdots  & \ddots & \vdots \\
    \rho_{q}       & \rho_{q}^2  & \dots &  \rho_{q}^m \\
\end{bmatrix}
\end{equation} and re-writing (\ref{eq:swarm_EOS}) in matrix form, we can obtain the virial coefficient matrix $A$ using the linear least squares approach, as follows:
\begin{equation}
\Psi = \Gamma A 
\end{equation}
\begin{equation}
A = (\Gamma^T \Gamma)^{-1}\Gamma^T \Psi
\end{equation}
where $\Gamma \in \mathbb{R}^{q\times m}$ is the density matrix, $\Psi \in \mathbb{R} ^{p\times q}$ is the residual values matrix, and $A \in \mathbb{R} ^{m\times p}$ is the coefficient matrix where each row of $A$ is associated with a virial coefficient and specifies how that coefficient varies as a function of swarm temperature $T_s$. In general, the accuracy of the equation of state increases if more number of virial coefficients $m$ that are included in the analysis. However, this approach is limited in practice by both quality and quantity of the simulation samples. In the presented work and Monte Carlo simulations, four residual terms were considered.

\section{Discussion}
In this section we discuss the tripartite relationship of macroscopic properties and the potential of using residuals to design specific task-oriented swarms.

    \begin{figure}[t]
    \begin{center}
    \includegraphics[width=0.7\linewidth]{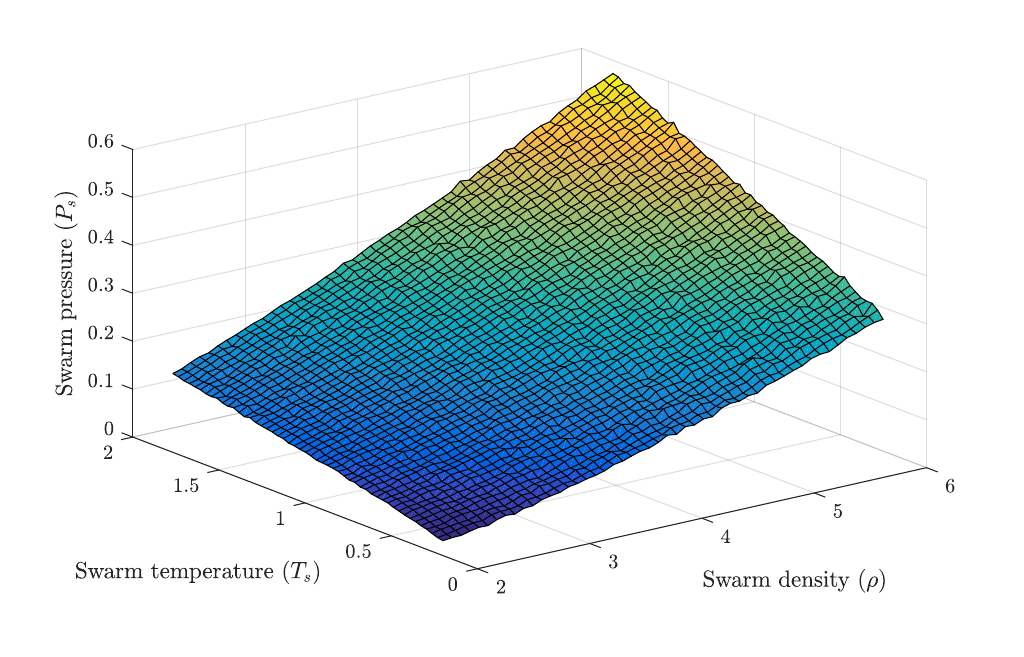}
    \caption{The relationship between swarm density, swarm temperature and swarm pressure, which manifests as a low-dimensional manifold in the three-dimensional $\rho-T_s-P_s$ space. Each of the 2500 simulations was run for 2000 time steps, with 50 samples for velocities ($0.2 < v < 1.8$) and 50 samples for densities ($2 < N/V < 5.6$).}
    \label{fig:pvt}
    \end{center}
    \end{figure} \textbf{}

\subsection{Tripartite Relation of the Macroscopic-scale Properties}
The previous sections introduced the macroscopic properties of swarm pressure and swarm temperature. Along with swarm density, these three properties may be used to describe the macroscopic state of a swarm, as for fluid systems. As shown in Fig. \ref{fig:pvt}, as swarm density and swarm temperature (which is proportional to agent velocity) increase, the swarm pressure increases as well. Specifically, within a fixed simulation area (i.e. constant `volume'), higher swarm density results in more `collisions' (or interactions) with the boundary constraints. Thus, these swarm behaviors mimic our knowledge of thermodynamics of real gases. Additionally, given two of the three macroscopic-scale variables for swarm systems enable us to identify the third variable using the low-dimensional manifold $h(P_s, T_s, \rho)=0$ shown in Fig. \ref{fig:pvt}. 

\subsection{Implications of the Cumulative Residual Term}
\begin{figure}[t]
    \begin{center}
    \includegraphics[width=0.7\linewidth]{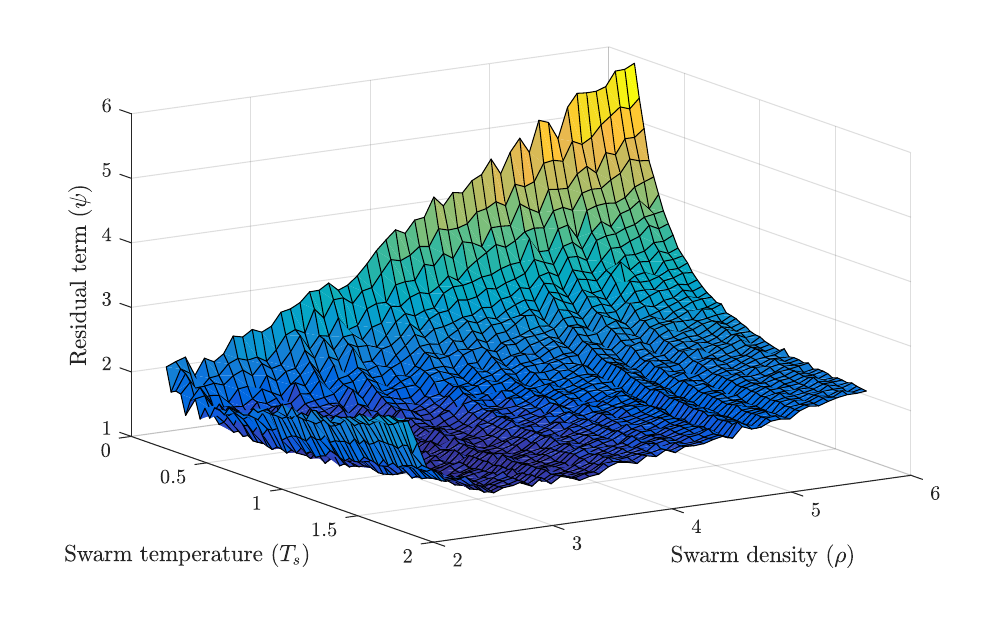}
    \caption{Unit-less cumulative residual term as a function of swarm density and temperature. As shown, for any given temperature $T_s$, the minimum residual term can be identified for a specific density in the corresponding $\rho-\psi$ plane. At this density, agents have the least interactions with each other, and mimic ideal-gas behavior most closely.}
    \label{fig:residual}
    \end{center}
    \end{figure}

Fig. \ref{fig:residual} shows the cumulative residual term as a function of swarm temperature and swarm density. The residual term indicates the deviation of the swarm from ideal-gas-like behavior, which also corresponds to deviations from non-interacting agent behavior. Knowledge of the residual term may be helpful depending on the context or task assignment of the swarm. For example, if the swarm is being tasked with patrolling or surveillance, it may be desirable to spread out the agent such that they cover large areas without significant interactions or redundancy of effort. In this scenario, swarm engineers may benefit from selecting macroscopic parameters such that the swarm behavior mimics ideal gases, i.e. has the smallest residual term. On the other hand, if the swarm has to perform collaborative tasks, such as building structures or mapping, then significant agent interactions may be required. Consequently, the swarm engineers would benefit from selecting macroscopic parameters that result in a large cumulative residual term, ensuring that agents interact repeatedly in their environment.

\section{Concluding remarks and future works}
Swarm engineering is currently making the transition from a research-centric endeavor to industrial applications, and its scalability is going to depend on being able to control large number of agents with relatively few control parameters. The presented thermodynamics-inspired macroscopic variables (swarm pressure, swarm temperature, and swarm density) offer a potential set of control parameters for a swarm governed by attractive-repulsive effects. The results also indicate that an empirical thermodynamics-inspired equation of state can yield a tripartite relationship between these macroscopic-scale properties, and that the relationship can be used to find one macroscopic variable given all others.

For unknown large-scale swarms, various valuable information (i.e. number of agents, agents absolute velocity and operational coverage) can be encoded in these macroscopic-scale properties (i.e. macrostate) of the system to provide a quantitative representation of the collective behavior of the swarm. Moreover, future work will leverage this information (along with the existing thermodynamics knowledge-base on phase transitions) to predict qualitative changes in swarm dynamics. Future work will also extend this approach to three-dimensional swarms, which also creates the possibility of studying macroscopic swarm dynamics in the context of equilibrium thermodynamic process, such as isobaric, isentropic, and isothermal processes, as well as subsequent expansion to non-equilibrium thermodynamic processes.

The presented work also does not fully address the relationship between the local interaction function and macroscopic-scale properties - the so-called micro-macro link. In future work, we intend to classify different collective behaviors of the swarm with limited macroscopic-scale data. It is worth mentioning here that for an unknown swarm, these macroscopic-scale properties are much easier to measure in comparison to estimating the microscopic-scale states of all agents. In fact, these issues were also faced by nascent thermodynamics researchers trying to analyze fluid behaviors in the 19\textsuperscript{th} century. Our future work will seek to leverage the significant advancements in thermodynamics over the past two centuries and re-purpose them to study large-scale swarms.

%Bibliography
\bibliographystyle{unsrt}  
\bibliography{ref}

\end{document}